\documentclass[fleqn,10pt]{wlscirep}
\usepackage[utf8]{inputenc}
\usepackage{float}
\usepackage[T1]{fontenc}
\usepackage{graphicx}
\usepackage{multirow}
\usepackage{xcolor}

\title{A Preliminary Investigation into Search and Matching for Tumour Discrimination in WHO Breast Taxonomy Using Deep Networks}

\author[1,2]{Abubakr Shafique}
\author[3]{Ricardo Gonzalez}
\author[4]{Liron Pantanowitz}
\author[5]{Puay Hoon Tan}
\author[6]{Alberto Machado}
\author[6]{Ian A Cree}
\author[1,2,*]{Hamid R. Tizhoosh}

\affil[1]{Rhazes Lab, Department of Artificial Intelligence and Informatics, Mayo Clinic, Rochester, MN, USA}
\affil[2]{Kimia Lab, University of Waterloo, Canada}
\affil[3]{Department of Laboratory Medicine and Pathology, Mayo Clinic, Rochester, MN, USA}
\affil[4]{Department of Pathology, University of Pittsburgh Medical Center, PA, USA}
\affil[5]{Luma Medical Centre, Singapore}
\affil[6]{International Agency for Research on Cancer, Lyon, France}

\affil[*]{corresponding author: tizhoosh.hamid@mayo.edu}

\keywords{World Health Organization Atlas, Breast Cancer, Image Search}

\begin{abstract}
Breast cancer is one of the most common cancers affecting women worldwide. They include a group of malignant neoplasms with a variety of biological, clinical, and histopathological characteristics. There are more than 35 different histological forms of breast lesions that can be classified and diagnosed histologically according to cell morphology, growth, and architecture patterns. 
Recently, deep learning, in the field of artificial intelligence, has drawn a lot of attention for the computerized representation of medical images. Searchable digital atlases can provide pathologists with patch matching tools allowing them to search among evidently diagnosed and treated archival cases, a technology that may be regarded as \emph{computational second opinion}. In this study, we indexed and analyzed the WHO breast taxonomy (Classification of Tumours 5th Ed.) spanning 35 tumour types. We visualized all tumour types using deep features extracted from a state-of-the-art deep learning model, pre-trained on millions of diagnostic histopathology images from the TCGA repository. Furthermore, we test the concept of a digital ``atlas'' as a reference for search and matching with rare test cases. The patch similarity search within the WHO breast taxonomy data reached over 88\% accuracy when validating through ``majority vote'' and more than 91\% accuracy when validating using top-$n$ tumour types. These results show for the first time that complex relationships among common and rare breast lesions can be investigated using an indexed digital archive.
\end{abstract}
\begin{document}

\flushbottom
\maketitle

\section{Introduction} \label{S-Introduction} 

The computerized calculations for a machine to emulate or even surpass human capabilities for performing a specific task are commonly referred to as \emph{artificial intelligence} (AI)~\cite{acs2020next}. A physicians' ability to identify patterns in data, such as images, and interpret those patterns in the context of other patient information as a whole is frequently a key component of diagnostic specialties such as radiology and diagnostic histopathology~\cite{van2021deep}. Studies have shown that AI is capable of performing at least as well as human experts on a number of medical image processing tasks~\cite{esteva2017dermatologist, de2018clinically, topol2019high,kalra2020pan}. Numerous fields of science, including those in health, are being rapidly transformed by AI and machine learning (ML)~\cite{koh2022artificial}. High-performance computers have recently emerged to expedite the training and deployment of AI models. Recent results from the application of AI to patient data in the field of medicine, and more particularly in diagnostic disciplines, are encouraging~\cite{yu2018artificial, litjens2017survey}. In digital diagnostic histopathology, pathologists perform meticulous visual examination of whole slide images (WSIs), a tedious task that is being increasingly supported by AI ~\cite{van2021deep}. Wider adoption of WSI analysis by the pathology community will facilitate  improving the diagnostic process~\cite{hamilton2014digital, cheng2022artificial}. Pathologists are increasingly using digital pathology (DP), which logically paves the way for computational pathology~\cite{kalra2020pan}. The gold standard for many diseases is the visual inspection of tissue morphology on Haematoxylin \& Eosin (H\&E) stained glass slides~\cite{shafique2021automatic}. Hence, we will focus on H\&E images in this study.  

\begin{figure}[ht]
\centering
\includegraphics[width=\linewidth,keepaspectratio]{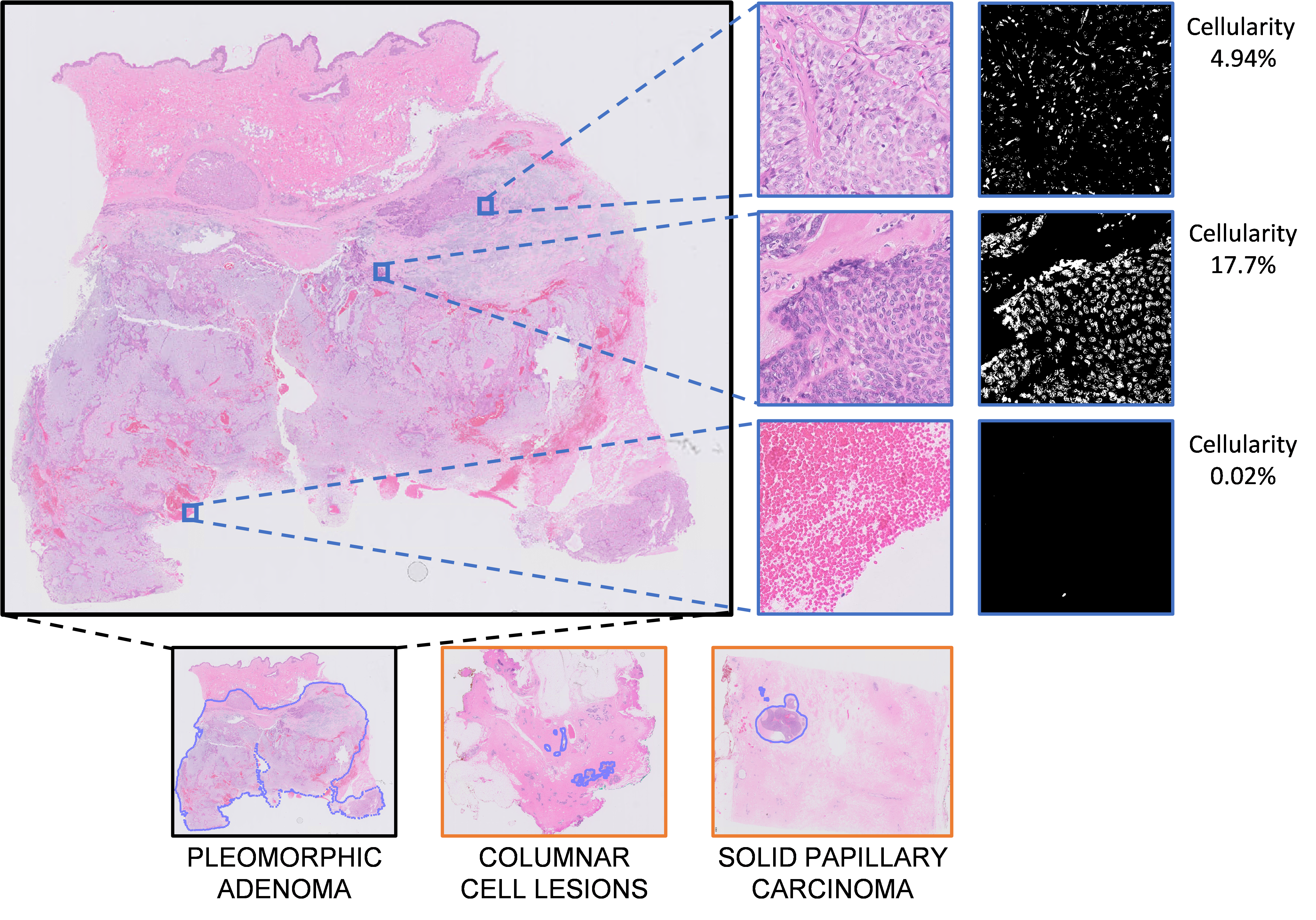}
\caption{Sample WSIs annotated by a pathologist (bottom row) from which patches can be extracted (top left). Sample patches based on  cellularity ratio (right panel) are also illustrated.}
\label{fig:WSI}
\end{figure}

In high income countries, breast cancer is the most prevalent cause of morbidity and also the second leading cause of death from cancer~\cite{weigelt2005breast, bray2018global}. Annually, there are 2.3 million new cases of breast cancer worldwide, accounting for 1 in 8 cancer diagnoses~\cite{arnold2022current}. Although mammography is a widely-used radiology screening tool, it is still difficult to interpret mammograms ~\cite{elmore2009variability}. Even the decisions of expert physicians may exhibit variability, as accuracy in the detection of cancer varies greatly among professionals~\cite{lehman2015diagnostic}. An accurate and timely diagnosis is critical for reliable prognosis and treatment planning, which can significantly increase patient survival rate~\cite{chen2020non, mckinney2020international}. Early-stage diagnosis 5-year survival is on average 91\%, while late-stage diagnosis 5-year survival is on average 26\%~\cite{chen2020non}. Chemotherapy raises the 15-year survival rate for breast cancer patients who are under 50 years of age by 10\%; for older patients, the increase is 3\%~\cite{weigelt2005breast, early2005effects}. The majority of patients with breast cancer have invasive breast (ductal) carcinoma of no special type (50\%–75\% of patients), followed by invasive lobular carcinoma (5\%–15\% of patients), with the remaining cases having mixed ductal/lobular carcinomas and other uncommon or rare histologies~\cite{waks2019breast}.

Breast tumour is a collective term for a number of malignancies with differing biological and pathological traits, clinical presentations, therapeutic strategies, clinical behaviors, and prognoses~\cite{mego2010molecular}. According to patterns of cell morphology, tumour growth, and architecture, there are more than 35 distinct histological types of breast tumours~\cite{dieci2014rare}. The most prevalent forms of breast lesions are invasive ductal breast carcinoma of no special type and invasive lobular carcinomas, although there are also a number of rare tumour types~\cite{arps2013invasive}. The biology and behavior of uncommon tumours can differ, necessitating potential targeted therapeutic modalities~\cite{dieci2014rare}. Several uncommon forms, such as \emph{encapsulated papillary carcinoma} and \emph{solid papillary carcinoma}, have distinctive symptoms and clinical manifestations. Some uncommon tumours like \emph{metaplastic carcinoma} and \emph{neuroendocrine} tumors are less well-understood and may have more unpredictable outcomes. In order to customize treatment strategies and enhance outcomes, it is crucial for afflicted patients that healthcare professionals are aware of the numerous forms of breast tumours, including the rare ones. As insights from research are rapidly incorporated into clinical practice, the classification of breast tumors continues to evolve~\cite{WHO2019breast}. Core biopsy, along with ancillary studies such as immunohistochemistry and sometimes molecular testing, assist with rendering an accurate diagnosis and are widely used~\cite{WHO2019breast}. 

In this paper, the digital data of World Health Organization (WHO) from its breast tumour classification book (5th Ed. 2019) with 35 different tumour types (containing both common and rare lesions) are processed, analyzed and visualized using state-of-the-art deep learning models. For the first time, this study introduced the t-distributed stochastic neighbor embedding (t-SNE) method as a visualization approach to assist cluster analysis for the entire WHO breast taxonomy data using deep features. This article is organized as follows: Data acquisition, preparation, and deep model processing are described in section~\ref{S-Methods}; Results and analysis are presented in section~\ref{S-Results}; Section~\ref{S-Discussion} includes the discussion and the conclusion.

\begin{table}[H]
\centering
\caption{\label{tab:CancerTypes}List of 35 breast tumours as primary diagnoses from the WHO classification of breast tumours ~\cite{WHO2019breast} used in our atlas as a reference for visualization and initial search and matching. Sample patches for all 35 breast tumours can be seen in Fig.~\ref{fig:All}.}
\begin{tabular}{|l|l|}
\hline
\textbf{No}. &  \textbf{Breast Tumours} \\
\hline
00 & Acinic Cell Carcinoma  \\
\hline
01 & Adenoid Cystic Carcinoma \\
\hline
02 & Adenomyoepithelioma \\
\hline
03 & Apocrine Adenosis and Adenoma  \\
\hline
04 & Atypical Ductal Hyperplasia  \\
\hline
05 & Atypical Lobular Hyperplasia  \\
\hline
06 & Carcinoma with Apocrine Differentiation  \\
\hline
07 & Columnar Cell Lesions, Including Flat Epithelial Atypia  \\
\hline
08 & Cribriform Carcinoma  \\
\hline
09 & Ductal Adenoma  \\
\hline
10 & Ductal Carcinoma in Situ  \\
\hline
11 & Encapsulated Papillary Carcinoma  \\
\hline
12 & Intraductal Papilloma  \\
\hline
13 & Invasive Breast Carcinoma of No Special Type  \\
\hline
14 & Invasive Lobular Carcinoma  \\
\hline
15 & Invasive micropapillary Carcinoma  \\
\hline
16 & Invasive Papillary Carcinoma  \\
\hline
17 & Lactating Adenoma  \\
\hline
18 & Lobular Carcinoma in Situ  \\
\hline
19 & Malignant Adenomyoepithelioma  \\
\hline
20 & Metplastic Carcinoma  \\
\hline
21 & Microglandular Adenosis  \\
\hline
22 & Microinvasive Carcinoma  \\
\hline
23 & Mucinous Carcinoma  \\
\hline
24 & Mucinous Cystadenocarcinoma  \\
\hline
25 & Neuroendocrine Carcinoma  \\
\hline
26 & Neuroendocrine Tumor  \\
\hline
27 & Papillary Ductal Carcinoma in Situ  \\
\hline
28 & Pleomorphic Adenoma  \\
\hline
29 & Radial Scar $/$ Complex Sclerosing Lesion  \\
\hline
30 & Sclerosing Adenosis  \\
\hline
31 & Secretory Carcinoma  \\
\hline
32 & Solid Papillary Carcinoma (in Situ and Invasive)  \\
\hline
33 & Tubular Adenoma  \\
\hline
34 & Tubular Carcinoma  \\
\hline
\end{tabular}
\end{table}

\begin{figure}[ht]
\centering
\includegraphics[width=\linewidth,keepaspectratio]{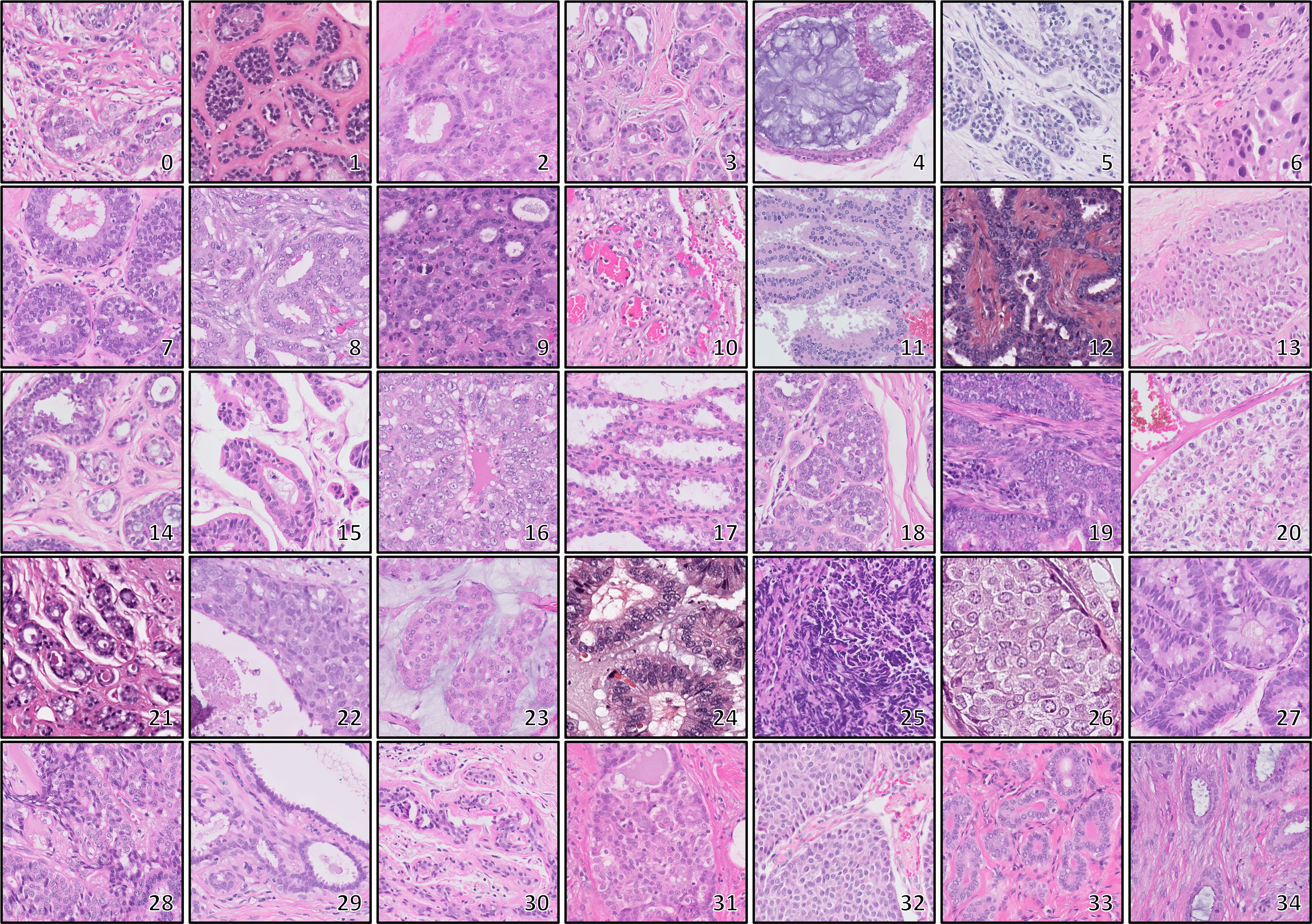}
\caption{Sample image patches extracted from annotated regions for 35 breast tumour types (see  diagnoses listed in Table.~\ref{tab:CancerTypes}).}
\label{fig:All}
\end{figure}

\section{Methods} \label{S-Methods}
\subsection{Dataset Preparation \& Acquisition} \label{S-DatasetAcquisition}
The data acquisition and preparation methods are described first, including annotations to focus on breast lesions when extracting features. Subsequently, we explain image representation via deep models, particularly KimiaNet that has been developed primarily for histopathology images using the TCGA (The Cancer Genome Atlas) repository.

\subsubsection*{Preparation}
WSIs are gigapixel digital image files that exhibit extremely large dimensions (generally containing millions of pixels). At the highest resolution and for large tissue sections scanned on slides, WSIs can exceed billions of pixels which may accordingly exceed the hardware limitations for necessary  computations~\cite{shafique2021automatic, tizhoosh2018artificial, srinidhi2021deep}. For many computational pathology applications, working with a WSI remains a challenge mainly because of its gigapixel size~\cite{tizhoosh2018artificial, lin2019fast}. Due to these factors, proper patch selection is the main emphasis of most computational techniques~\cite{riasatian2021fine, adnan2020representation, chenni2019patch}. An optimal patching algorithm serves as the ``Divide'' in a \emph{Divide \& Conquer} paradigm to solve a complex problem. Hence, patching should select a rather small number of patches (for efficiency), but should not miss any relevant patches (to ensure correct diagnosis).

In our experiments, patches of size $512$ $\times$ $512$ were densely extracted (with 20\% to 80\% patch overlap) from the WSIs within the malignant regions (see fig.~\ref{fig:WSI}). The extent of overlap between patches is determined by the area of the tissue that has been delineated. If the delineated area is small, there is more overlap between patches to increase the number of patches. Conversely, if the area is larger, the overlap between patches is smaller, hence generating a smaller number of patches. The delineated regions may contain healthy/benign patches which would then be falsely indexed as malignant patches and affect reliable information retrieval. To avoid additional visual inspections, \emph{low cellularity} can be used as a filter to identify patches that are more likely to be benign. Some cancers like carcinomas may typically be characterized by excessive cell growth, which is mostly manifested in \emph{hypercellular} patches~\cite{travis2014pathology, riasatian2021fine}.
We calculated the cellularity of each patch by using H\&E color deconvolution from RGB to H\&E channels~\cite{scikit-image, ruifrok2001quantification}. The cellularity ratio for each patch was then determined by creating a binary mask from the hematoxylin channel using an empirically determined constant threshold (see Fig.~\ref{fig:WSI}). The number of white pixels in the produced mask over the total number of patch pixels approximates the "cellularity" in this context.Figure.~\ref{fig:WSI} shows a few samples of tumor with different cellularity levels, and patches with cellularity above 8\% are selected as abnormal patches and other patches are discarded as healthy/benign. 

\begin{figure}[ht]
\centering
\includegraphics[width=\linewidth,keepaspectratio]
{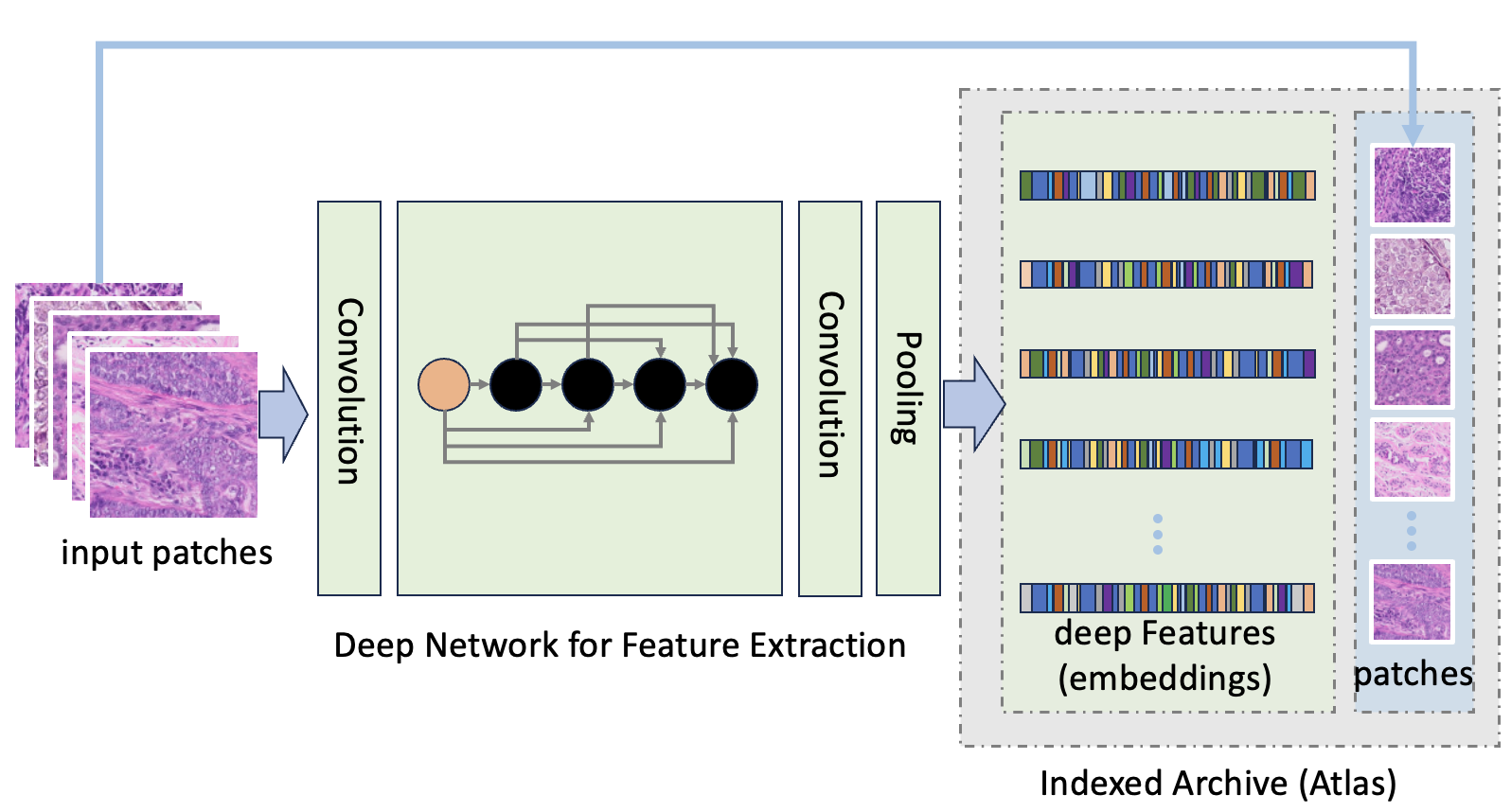}
\caption{Schematic creation of a breast atlas: Feature vectors (or embeddings) of the input patches generated using a deep network (\emph{KimiaNet}~\cite{riasatian2021fine}) consisting of four dense blocks and consecutive use of operations ``convolution'' (i.e., filtering) and ``Pooling'' (i.e., sub-sampling). The patches together with the embeddings (i.e., deep features) build the indexed archive or atlas.}
\label{fig:KIMIANet}
\end{figure}

\subsubsection*{Acquisition \& Annotation}
The Department of Anatomical Pathology, Singapore General Hospital, provided anonymized archival digital slides from the time period 2011 to 2017. 
Using the Philips Ultra Fast Scanner (UFS), the selected slides were converted into WSIs using the Philips Intellisite Pathology System (PIPS). 
WSIs were subsequently downloaded from PIPS in iSyntax format. In this initial study, 38 WSIs from different tumours were used, containing 35 lesion types, including some rare cases (see Table.~\ref{tab:CancerTypes} and Fig.~\ref{fig:All})~\cite{cheng2022artificial}. From a total of 38 WSIs, 35 distinct WSIs (constituting 21,847 high cellular patches filtered from 32,530 total patches) represented our digital atlas using data from the WHO classification of breast tumours book~\cite{WHO2019breast}. From the limited available data, the remaining three WSIs (2,406 high cellular patches with the following primary diagnoses: 11, 12, and 32 from Table.~\ref{tab:CancerTypes}) represented the challenging test cases for search and matching against our digital atlas. 

The Automated Slide Analysis Platform (ASAP) (developed by the Computational Pathology Group of the Diagnostic Image Analysis Group at the Radboud University Medical Center)~\cite{ASAP} was used for annotations. All lesions were finely delineated with the polygon tool using high magnifications by a pathologist (R.G.) involved in the study. The specific magnification factors varied among different lesions and fields of view according to the degree of irregularity along the lesion borders. For non-circumscribed lesions (e.g., radial scars), the annotation boundaries were arbitrarily selected to include only relevant epithelial and/or stromal tissue components (some annotated examples are shown in Fig.~\ref{fig:WSI}). All annotated contours were exported to XML format to generate masks for patch selection/extraction (see Fig.~\ref{fig:All}). 

\subsection{Deep Models for Image Representation} \label{S-DeepNetwork}

Deep neural networks (DNNs) have supplanted other machine learning (ML) techniques for the recognition of visual objects~\cite{huang2017densely,yu2021convolutional}. DNNs have the intrinsic ability to learn complicated characteristics directly from image data, which is a significant advantage over conventional (handcrafted) computer vision algorithms~\cite {anwar2018medical, Alsaafin2023}. The use of trained deep models enables us to define intelligent software to assist pathologists in diagnostic and research purposes. Whereas DNNs may be employed to make a decision, \emph{representation} of diagnostic histopathology images through ``deep features'' is a pivotal aspect of DNNs. Various factors such as color, cellular density, gland size, tissue textures, magnification levels, and other factors may be captured in deep image representation~\cite{slidders1981study}. Diagnostic histopathology images are represented by feature vectors, often called \emph{embeddings}, a very compact representation that can be used for a number of tasks such as identification, segmentation, categorization, clustering, as well as search and matching. Unstructured data can be represented by such embeddings in a $d$-dimensional feature space enabling visual inspections in novel ways.

In our investigation, we used KimiaNet, a DenseNet-121 architecture  specialized and trained with over 240,000 histopathology patches to distinguish 30 tumour types including normal tissue, benign tumors, and various types of malignant tumors.~\cite{riasatian2021fine}. This architecture is known for its ability to extract rich feature representations from images with densely connected layers. DeneseNet weights, the starting point for training KimiaNet,  were trained using 1.2 million natural images to separate 1000 classes~\cite{huang2017densely}. Adapting DenseNet to histopathological images through fine-tuning was crucial in order to account for their unique features and characteristics. 

The densely extracted patches from WHO based WSIs were fed into the KimiaNet to generate embeddings from the last pooling layer (see Fig.~\ref{fig:KIMIANet}). The feature vector of length 1024 acquired from KimiaNet represents the tissue morphology of the patch and hence can be stored as the ``index'' of that tissue fragment in a digital atlas. Furthermore, principal component analysis (PCA) was used to reduce the dimensionality and pick the top-50 principal components in order to remove the noise and redundancy~\cite{chen2021lung}. Finally, these representations (full-length vector and the top-50 principal components) can be used for cluster analysis and t-distributed stochastic neighbor embedding (t-SNE) visualization~\cite{scikit-learn, van2008visualizing}. 

\section{Results} \label{S-Results}

In this investigation, 35 types of breast tumours were analyzed and visualized by indexing images through a specific type of deep model called KimiaNet. Moreover, we introduce the notion of a digital breast tumour atlas that encompasses reference images of all breast lesions. Such an atlas can be employed for visualization as well as for matching new (undiagnosed) patients with standard cases in the WHO classification of breast tumours book~\cite{WHO2019breast}. We performed the experiments on a Dell Desktop with 1x Intel(R) Core(TM) i9-7900X CPU (10 cores, 3.30GHz), 1x Nvidia GeForce GTX 1080 (v-RAM 8 GB), and 64 GB RAM.

\subsubsection*{Search \& Matching}
Verifying and validating histologic similarity is a challenging task. The ideal setting for a comprehensive validation would require matching many patients from different hospitals and visual inspection by multiple pathologists over a period of time. To quantify the performance of the search task in this preliminary research, we treated search and matching \emph{like a classifier} to simplify things. One of the main advantages of employing classification methods is that they are simple to validate; every image can be classified as either belonging to a class or not, which is a binary concept that can be quantified by counting the misclassified instances. However, the concept of resemblance in image search is fundamentally a gradual concept (i.e., in many situations, the question of similarity may not be answered with a straightforward yes/no) and primarily a matter of degree (\emph{very similar}, \emph{quite dissimilar}, etc.). In addition, a distance measure is typically used to determine how similar (or dissimilar) two images are, a measurement that uses metrics like  Euclidean distance that quantifies the dissimilarity between two given feature vectors representing the two images. The classification-based tests we use may be overly conservative to evaluate the search results and blind to anatomical commonalities across many tumour types.

\begin{table}[]
\centering
\caption{\label{tab:classification} Precision, recall, F1-score, and accuracy of top-1 (Majority-1) search and matching results when treated as classifier.}
\begin{tabular}{|l|c|c|c|c|}
\hline
\textbf{Breast Diagnosis}    & \textbf{Precision} & \textbf{Recall} & \textbf{F1-score} & \textbf{No. of Patches} \\ \hline
11 (Encapsulated Papillary Carcinoma)       & 0.96      & 0.95   & 0.95     & 951     \\ \hline
12 (Intraductal Papilloma)      & 1.0       & 1.0    & 1.0      & 981     \\ \hline
32 (Solid Papillary Carcinoma (in Situ and Invasive))      & 0.98      & 0.47   & 0.64     & 474  
 \\ \hline
\textbf{Accuracy} &           &        & 0.88     & 2406    \\ \hline
\end{tabular}
\end{table}

\begin{figure}[ht]
\centering
\includegraphics[width=0.75\textwidth]{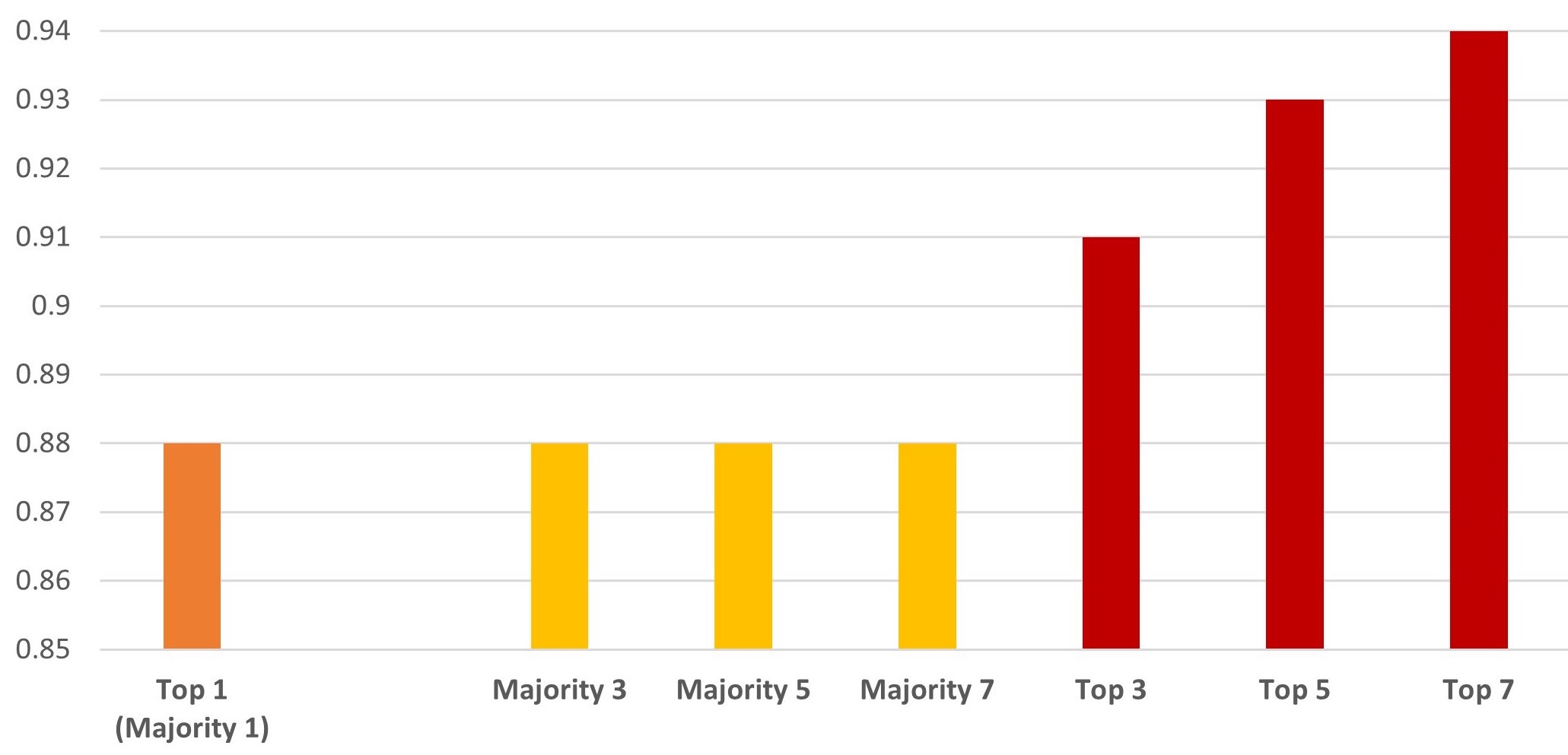}
\caption{The overall accuracy of the breast tumour atlas search and matching results are shown according to the Top-$n$ and Majority-$n$ validation approaches. Top-1 and Majority-1 are practically the same, and for this reason, they are shown as one bar.}
\label{fig:Accuracy}
\end{figure}

In this study, we used the $k$-nearest neighbors ($k$-NN) algorithm to find the top-matched (most similar) embeddings when searching through the atlas to find the best matches. We can then retrieve the top-matched images that meet the search criteria (nearest neighbors based on the Euclidean distance). For all the distance metrics calculations in search and matching, full-length feature vectors of the extracted patches are used. Computer vision literature focuses on top-$n$ accuracy (if any one of the $n$ search results is correct, the search is considered to be successful). However, we use ``majority-$n$ accuracy'' which is a much more reliable validation scheme for medical imaging; only if the majority among the $n$ search results were correct, the search considered correct (see Fig.~\ref{fig:Accuracy}, in our case $n$ = 1, 3, 5, and 7). 

\begin{figure}[htb]
\centering
\includegraphics[width=\linewidth,keepaspectratio]{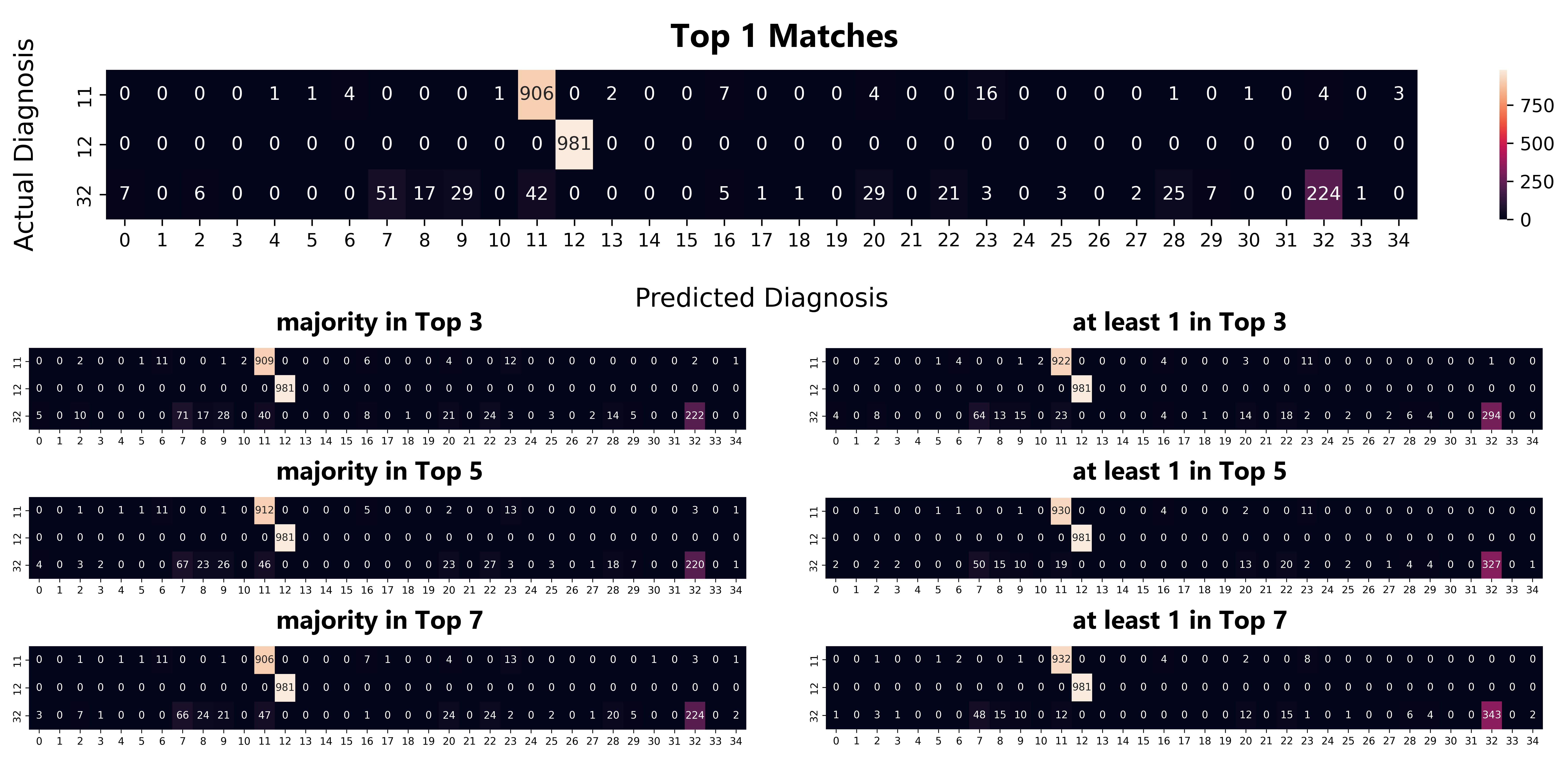}
\caption{Confusion matrices for the classification of primary diagnoses using Top-$n$ and Majority-$n$ validation approaches (in our case $n$ = 1, 3, 5, and 7). Top-1 and Majority-1 are the same, and for this reason, they are shown as one confusion matrix.}
\label{fig:CM}
\end{figure}

Table.~\ref{tab:classification} shows the precision, recall, F1-score, and accuracy of top-1 search results when treated as a classifier. Furthermore, Fig.~\ref{fig:Accuracy} shows the overall accuracy results using both majority-$n$ and top-$n$ approaches. Furthermore, looking at the confusion matrix is another method for evaluating the effectiveness of the search (see Fig.~\ref{fig:CM}). The relative frequency of each subtype among the top-$n$ and majority-$n$ search results for a specific subtype can be used to determine the values needed to create the confusion matrices.

\begin{figure}
\centering
\includegraphics[width=0.8\linewidth,keepaspectratio]{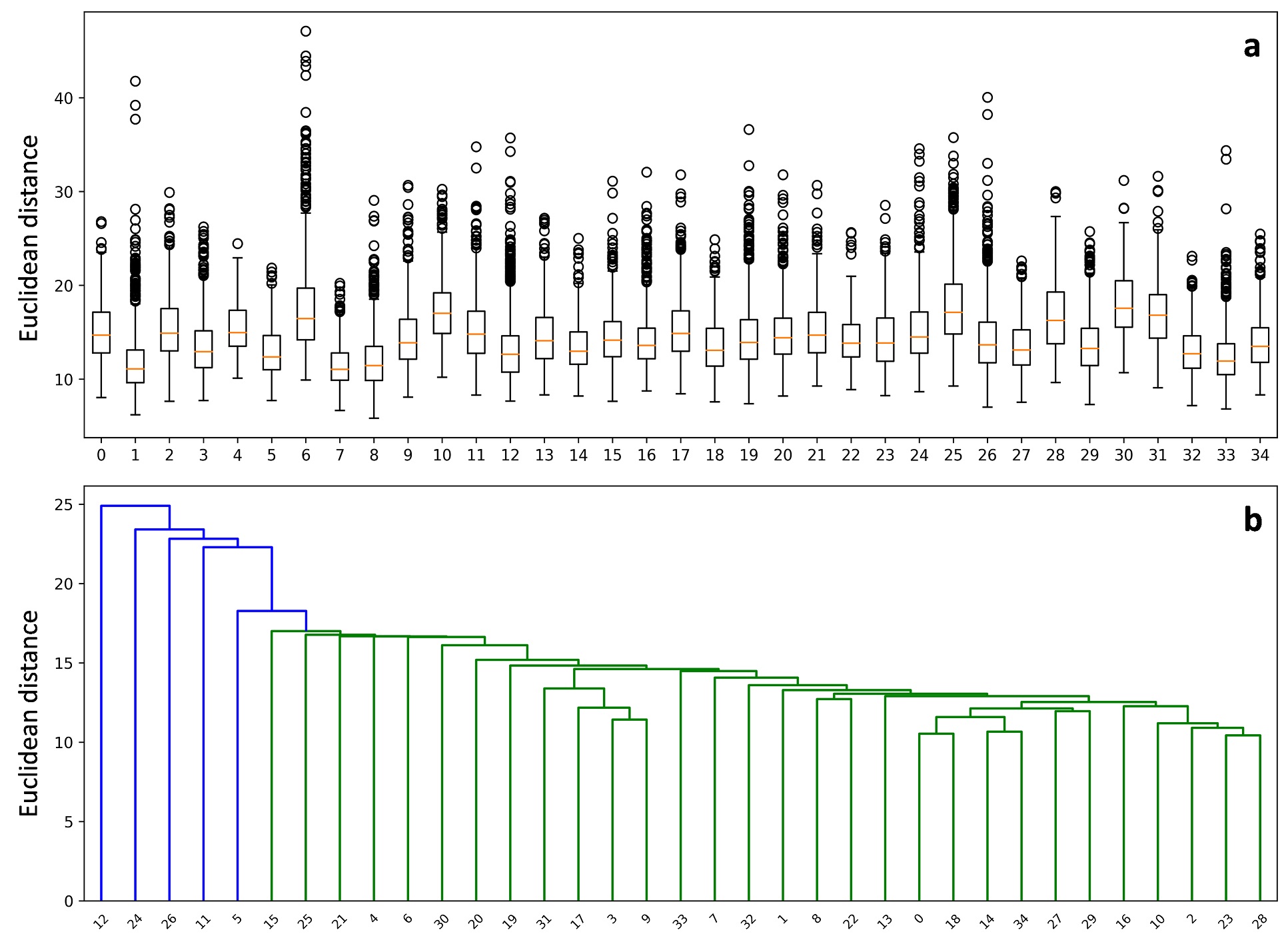}
\caption{Euclidean distance-based inter- and intra-cluster analysis of feature vectors from 35 types of breast tumour (see Table.~\ref{tab:CancerTypes}): (a) shows the intra-class distance variation using boxplot, (b) the constructed dendrogram shows the hierarchical clustering single linkage using 35 centroids.}
\label{fig:frequency}
\end{figure}

\textbf{Top Frequent tumour types --} The likely tumour type based on the most frequent matches, that we call ``top-3@top-n'', can be seen in Table.~\ref{tab:frequency}. Performing this match means we find the top 3 most frequent tumour types among the top $n$ search results. One of the significant advantages of the top suggestions based on the frequently matched samples narrows down the diagnosis of rare cases such as solid papillary carcinoma (32 from Table.~\ref{tab:CancerTypes}). Table.~\ref{tab:frequency} shows that solid papillary carcinoma is most likely similar to itself, and also similar to columnar cell lesions including flat epithelial atypia and encapsulated papillary carcinoma (class 32, 7, and 11 from Table.~\ref{tab:CancerTypes}) in the respective order based on the matching frequency for majority-$k$. These lesions indeed all have similar histologic features, including papillary architecture, columnar cells, and solid areas~\cite{nassar2006clinicopathologic}.

\begin{table}
\centering
\caption{\label{tab:frequency} Top 3 most frequent search and matching results at majority-$k$ (with $k=1, 3, 5, 7$) for $m$ patches of a test WSI. Additionally, this table shows the confidence of matching results based on the percentage of matching degrees.}
\resizebox{\textwidth}{!}{\begin{tabular}{|l|c|ccc|ccc|ccc|ccc|}
\hline
& & \multicolumn{3}{c|}{\textbf{\begin{tabular}[c]{@{}c@{}}Most Frequent Class\\  at Majority $K$ = 1\\ (Class / Frequency)\end{tabular}}} & \multicolumn{3}{c|}{\textbf{\begin{tabular}[c]{@{}c@{}}Most Frequent Class\\  at Majority $K$ = 3\\ (Class / Frequency)\end{tabular}}} & \multicolumn{3}{c|}{\textbf{\begin{tabular}[c]{@{}c@{}}Most Frequent Class\\  at Majority $K$ = 5\\ (Class / Frequency)\end{tabular}}} & \multicolumn{3}{c|}{\textbf{\begin{tabular}[c]{@{}c@{}}Most Frequent Class\\  at Majority $K$ = 7\\ (Class / Frequency)\end{tabular}}} \\ \hline
\multicolumn{1}{|c|}{\textbf{Test Class}}                                           & \textbf{Patches}     & \multicolumn{1}{c|}{\textbf{First}}               & \multicolumn{1}{c|}{\textbf{Second}}               & \textbf{Third}              & \multicolumn{1}{c|}{\textbf{First}}               & \multicolumn{1}{c|}{\textbf{Second}}               & \textbf{Third}              & \multicolumn{1}{c|}{\textbf{First}}               & \multicolumn{1}{c|}{\textbf{Second}}               & \textbf{Third}              & \multicolumn{1}{c|}{\textbf{First}}               & \multicolumn{1}{c|}{\textbf{Second}}               & \textbf{Third}              \\ \hline
\multicolumn{1}{|c|}{\textbf{11}}  & \multirow{2}{*}{951} & \multicolumn{1}{c|}{11 / 906}& \multicolumn{1}{c|}{23 / 16} & 16 / 7 & \multicolumn{1}{c|}{11 / 909}                     & \multicolumn{1}{c|}{23 / 12} & 6 / 11  & \multicolumn{1}{c|}{11 / 912} & \multicolumn{1}{c|}{23 / 13} & 6 / 11  & \multicolumn{1}{c|}{11 / 906} & \multicolumn{1}{c|}{23 / 13}  & 6 / 11  \\

\textbf{\begin{tabular}[c]{@{}l@{}}Percentage of\\ total patches (\%)\end{tabular}} & & \multicolumn{1}{c|}{95.27} & \multicolumn{1}{c|}{1.68} & 0.73 & \multicolumn{1}{c|}{95.58} & \multicolumn{1}{c|}{1.26}  & 1.15 & \multicolumn{1}{c|}{95.89} & \multicolumn{1}{c|}{1.36} & 1.15  & \multicolumn{1}{c|}{95.27}  & \multicolumn{1}{c|}{1.82} & 1.15   \\ \hline

\multicolumn{1}{|c|}{\textbf{12}}  & \multirow{2}{*}{981} & \multicolumn{1}{c|}{12 / 981}                     & \multicolumn{1}{c|}{-}                             & -                           & \multicolumn{1}{c|}{12 / 981}                     & \multicolumn{1}{c|}{-}                             & -                           & \multicolumn{1}{c|}{12 / 981}                     & \multicolumn{1}{c|}{-}                             & -                           & \multicolumn{1}{c|}{12 / 981}                     & \multicolumn{1}{c|}{-}                             & -                           \\
\textbf{\begin{tabular}[c]{@{}l@{}}Percentage of\\ total patches (\%)\end{tabular}} &                      & \multicolumn{1}{c|}{100.0}                        & \multicolumn{1}{c|}{-}                             & -                           & \multicolumn{1}{c|}{100.0}                        & \multicolumn{1}{c|}{-}                             & -                           & \multicolumn{1}{c|}{100.0}                        & \multicolumn{1}{c|}{-}                             & -                           & \multicolumn{1}{c|}{100.0}                        & \multicolumn{1}{c|}{-}                             & -                           \\ \hline

\multicolumn{1}{|c|}{\textbf{32}}                                                   & \multirow{2}{*}{474} & \multicolumn{1}{c|}{32 / 224} & \multicolumn{1}{c|}{7 / 51} & 11 / 42 & \multicolumn{1}{c|}{32 / 222}& \multicolumn{1}{c|}{7 / 71} & 11 / 40  & \multicolumn{1}{c|}{32 / 220}   & \multicolumn{1}{c|}{7 / 67}  & 11 / 46 & \multicolumn{1}{c|}{32 / 224} & \multicolumn{1}{c|}{7 / 66}  & 11 / 47 \\

\textbf{\begin{tabular}[c]{@{}l@{}}Percentage of\\ total patches (\%)\end{tabular}} && \multicolumn{1}{c|}{47.26} & \multicolumn{1}{c|}{10.76} & 8.86   & \multicolumn{1}{c|}{46.83}  & \multicolumn{1}{c|}{14.98}& 8.43 & \multicolumn{1}{c|}{46.41} & \multicolumn{1}{c|}{14.13} & 9.70 & \multicolumn{1}{c|}{47.26} & \multicolumn{1}{c|}{13.92} & 9.91 \\ \hline
\end{tabular}}
\end{table}

\subsection*{Feature Analysis} \label{S-FeatureAnalysis}
Understanding spatial and temporal patterns can be aided by using cluster analysis. Data patterns and anatomic groupings are detected using unsupervised cluster analysis when there is a higher degree of similarity between the points in one cluster than between the points in another cluster. Identification of meaningful and comprehensible groups in the data, which can help reveal underlying patterns and structures, is the major objective of cluster analysis~\cite{kenny2022identification, karim2021deep}. Also, clustering is crucial for offering a more explainable and understandable representation of complex data sets~\cite{karim2021deep, oyelade2016clustering}. Additionally, cluster analysis can identify connections and linkages between data points that may not be obvious in the raw data. 

In this study, for the first time, we performed Euclidean distance-based inter- and intra-class analysis of deep feature vectors for all breast lesions listed in the WHO classification of breast tumours Book~\cite{WHO2019breast}. We used k-means clustering to determine the \emph{centroids} (i.e., the typical instance for each tumour type). The Euclidean distance of each point (i.e., each patch represented by its feature vector) from the centroid for each class. Subsequently, we can analyze the standard deviation of the distance within each tumour type as shown in Fig.~\ref{fig:frequency} (a). Furthermore, for inter-cluster analysis, centroids are used to calculate the single linkage among tumour types, which is also known as hierarchical clustering~\cite{shaffer1979single, landau2011cluster, mohbey2013experimental} (see Fig.~\ref{fig:frequency} (b)). Cluster linkage aims to build a hierarchical representation of the data, with dissimilar data points being divided into different clusters at higher levels of the hierarchy and comparable data points being grouped together at lower levels. The hierarchical clustering single linkage algorithm starts with each data point as a separate cluster, and then iteratively merges the closest pairs of clusters into larger clusters until all points belong to a single cluster~\cite{ros2019hierarchical}. Here, we used this method to identify meaningful groupings and the nearest similar subtype (see Fig.~\ref{fig:frequency} (b)).

\begin{figure}
\centering
\includegraphics[width=\linewidth,keepaspectratio]{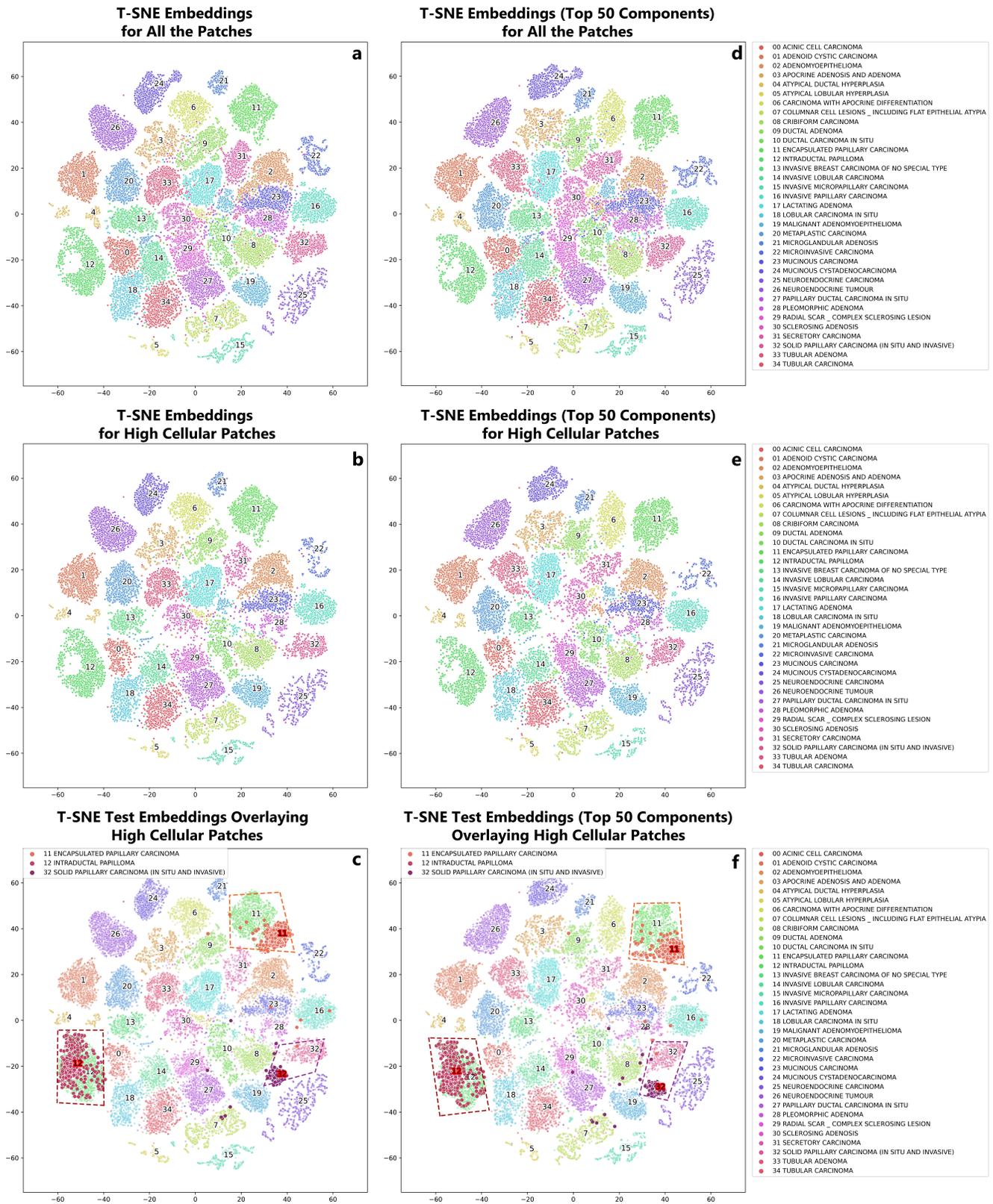}
\caption{t-SNE~\cite{van2008visualizing} visualization of patches using KimiaNet: (a) \& (d) the embeddings of all atlas patches extracted from 35 distinct WSIs, (b) \& (e) the embeddings of high-cellular patches filtered based on the cellularity ratio, and (c) \& (f) the embeddings of high-cellular patches from the test WSIs overlayed on our high-cellular atlas (b) \& (e) (highlighted in red). The legend of 35 WSIs in our atlas is on the right side of (d), (e), and (f), whereas the legend of test WSIs is on the top left side of (c) \& (f).}
\label{fig:TSNE}
\end{figure}

Additionally, we also tried to retain the top-50 principal components of our 1024-length embeddings using PCA, from which we calculated the silhouette coefficient~\cite{rousseeuw1987silhouettes}, Davies-Bouldin score~\cite{davies1979cluster}, and Calinski-Harabasz score~\cite{calinski1974dendrite} using both types feature vectors (1024-length, and top 50 principal components) of all patches and their corresponding labels (see Table~\ref{tab:clusterAnalysis}). These three commonly used measures of cluster quality can be used to evaluate the features of the atlas (Table~\ref{tab:CancerTypes}). The silhouette coefficient ranges from $-1$ to $1$, with higher values indicating a better separation between clusters. With the full 1024-length vector the silhouette coefficient is $0.14$ but with the top-50 principal components of the vector, the silhouette coefficient improved slightly (which is $0.17$, see Table ~\ref{tab:clusterAnalysis}). Similarly, a Davies-Bouldin score for the two types of vectors was also calculated. A Davies-Bouldin score of $2.17$ was calculated for full-length vectors and the reduced vector scored $2.00$ (see Table~\ref{tab:clusterAnalysis}). Ideally, a Davies-Bouldin score closer to zero would indicate better clustering. The scores, we observed, indicate that choosing the top-50 principal embeddings marginally improves the clustering. In the end, the Calinski-Harabasz Index (also known as the Variance Ratio Criterion) was also calculated. For full-length vectors, a score of $666.79$, and for top-50 a score of $822.22$ for $35$ clusters was achieved (see Table~\ref{tab:clusterAnalysis}). These numbers suggest that the clustering is relatively good given the complexity of the data. However, clustering improved a little by removing the redundant components of the embeddings and just retaining the top-50.

\begin{table}[ht]
\centering
\caption{\label{tab:clusterAnalysis} Silhouette coefficient, Davies-Bouldin score, and Calinski-Harabasz score for the breast cancer atlas evaluation. These metrics suggest that the clustering for this complex dataset with 35 types of breast cancer is moderate in quality. However, the quality of clusters improved significantly when choosing the top-50 principal components.}
\begin{tabular}{|c|c|c|c|c|c|}
\hline
\textbf{Classes} & \textbf{Features} & \textbf{Feature Length} & \textbf{\begin{tabular}[c]{@{}c@{}}Silhouette \\ Coefficient\end{tabular}} & \textbf{\begin{tabular}[c]{@{}c@{}}Davies-Bouldin\\ Score\end{tabular}} & \textbf{\begin{tabular}[c]{@{}c@{}}Calinski-Harabasz\\ Score\end{tabular}} \\ \hline
35& 21,847 & 1024 & 0.14 & 2.17 & 666.79 \\ \hline
35& 21,847 & 50 & 0.17 & 2.00 & 822.22 \\ \hline
\end{tabular}
\end{table}

\subsubsection*{Breast Pan-Subtype Visualization}
For visualization  of high-dimensional data through scatter plots, the neighborhood-preserving multidimensional scaling method (short t-SNE) is frequently utilized. Here, we utilized t-SNE to visualize the patches of 35 subtypes (see Fig.~\ref{fig:TSNE} (a) to (e)).
t-SNE was used to generate the maps for two types of vectors (full 1024-length vector, and top-50 principal components from the full-length vector) to visually compare the difference which we observed in feature analysis. This is an actual visualization of the breast cancer atlas although every group is one prototypical patient represented through its densely sampled patches. In Fig.~\ref{fig:TSNE}(a) \& (d), we can see the embeddings of all 32,530 patches from the 35 primary diagnoses. Many benign or healthy patches that are mislabeled as malignant patches are seen here because there isn't a distinct boundary separating them. Often, cancer patches, particularly in the case of carcinomas, exhibit hypercellularity. For this reason, we filtered 21,847 high-cellular patches out to generate a more distinct visualization of subtypes in Fig.~\ref{fig:TSNE}(b) \& (e) where all the subtypes are more distinctly separated, a better reference for search and matching. 

In order to visually match the high-dimensional embeddings, we used the 2,406 patches from three test cases (WSIs) to plot their embeddings in 2D space and overlaid on our atlas (shown in Fig.~\ref{fig:TSNE}(c) \& (f)). The bounding boxes in Fig.~\ref{fig:TSNE}(c) \& (f) point to the embeddings of test cases and their corresponding matching primary diagnosis.

\section{Discussion \& Conclusion} \label{S-Discussion}
\subsection*{Discussion}
The concept of building and using an atlas is not a new idea as it dates back to ancient times. However, the modern concept of a \emph{digital atlas} as a well-curated collection of cases to be indexed by modern AI for search and matching is novel. In this paper, we have used well-curated WHO breast images to create a first iteration of a digital atlas as a reference for search and matching. This preliminary study aimed at analyzing tumour types through their vector representations from state-of-the-art deep networks  using cluster analysis and visualization techniques. The mere availability of such visualizations and mapping a new case/tumour into the atlas is a new venue for breast cancer researchers and could one day perhaps replace human expert consultation to help diagnose challenging breast cancer cases.

In our experiments, we successfully demonstrated that the concept of an atlas for  matching breast tumours may provide value especially when we examine the entirety of known tumour types. Our results show more than 88\% overall accuracy for 2,406 test patches matching against 21,847 patches from the atlas (see Fig.~\ref{fig:Accuracy}). Other than accuracy, we also analyze the matching results based on the majority vote and top-n accuracy (see Fig.~\ref{fig:CM}). In the literature, the medical community prefers to use the majority vote approach considering the sensitivity of the domain. Additionally, rare cases are less well understood and may have more unpredictable outcomes. For this reason, for the very first time, we have introduced the top k most likely matching primary diagnoses among top-n search results to narrow down the broad diagnosis observations based on the matching confidence (percentage of matching instances), which can be seen in Table.~\ref{tab:frequency}. 

Analyzing the WHO breast classification of breast tumours book~\cite{WHO2019breast} data with diverse lesion types (see Table.~\ref{tab:CancerTypes}) provides an opportunity to examine the complexity of the diseases for breast as a primary site and identify meaningful groupings and linkages among different primary diagnoses. Because of this, we applied Euclidean-based inter- and intra-cluster analysis on the feature vectors (see Fig.~\ref{fig:frequency}). From Fig.~\ref{fig:frequency}(a), we could see a high intra-cluster variation in ``Carcinoma with apocrine differentiation'' (subtype no. 6 according to Table.~\ref{tab:CancerTypes}) due to the ink artifact in the WSI which experts often use to identify and mark the resection margins during the gross examination of a tissue specimen. Carcinoma with apocrine differentiation, is a rare breast cancer type and accounts for less than 1\% of all breast cancer cases~\cite{vranic2017apocrine}. For the inter-cluster linkage (see Fig.~\ref{fig:frequency}), the centroid from each case (prototypical patch for a subtype) is used to provide insight into the nearest possible subtypes, which is very important when observing rare diseases. From Fig.~\ref{fig:frequency}(b), we could observe that ``Invasive lobular carcinoma'' and ``Tubular carcinoma'' (subtype no. 14, and 34 according to Table.~\ref{tab:CancerTypes}) have the shortest link with each other because they are both malignant, meaning that they have the potential to invade within the breast tissue and may metastasize to other parts of the body, though tubular carcinoma is an invasive breast cancer type with excellent prognosis. Invasive lobular carcinoma is a type of breast cancer that tends to infiltrate more widely than clinicoradiologically suspected as compared to invasive ductal carcinoma. It accounts for about 10-15\% of all breast cancers~\cite{mccart2021invasive}, whereas tubular carcinoma is characterized by the formation of small, tube-like structures and accounts for about 1-5\% of all breast cancers~\cite{min2013tubular}. On the other hand, ``Intraductal papilloma'' (subtype no. 12 according to Table.~\ref{tab:CancerTypes}) is distinct with the greatest distance from all other cases because it is a type of benign breast tumour that arises from the epithelial cells lining the ducts of the breast. It is a common breast abnormality, accounting for about 10\% of all breast biopsies~\cite{shouhed2012intraductal}. Intraductal papilloma is characterized by the growth of small, finger-like projections (papillae) within the breast ducts~\cite{ohuchi1984origin}. These papillae can be lined with cells that are similar in appearance to those found in the ducts of the breast and may contain a central core of fibrous tissue, blood vessels, and/or small ducts~\cite{wei2016papillary}. Intraductal papilloma can occur at any age but is most commonly seen in pre-menopausal women~\cite{wei2016papillary}. In addition, it is crucial to distinguish intraductal papillomas from intraductal carcinoma, a type of breast cancer that can have a similar appearance on mammography and biopsy. While intraductal papillomas are benign, intraductal carcinoma is a malignant lesion that requires prompt treatment.

Finally, we assess the patch embeddings statistically and visually using euclidean distance based cluster analysis and t-SNE dimensionality reduction technique. Full-length Feature for inter- and intra-cluster analysis can be seen in Fig.~\ref{fig:frequency}, Whereas, Table.~\ref{tab:clusterAnalysis} metrics shows the quality of 35 clusters. We also used PCA to choose the top-50 principal components, which were then compared with the full-length vectors to assess the quality of these class-based clusters. The quality of the top-50 principal components was slightly better than the full-length vector clusters as shown in Table.~\ref{tab:clusterAnalysis}. Full-length vectors and the top-50 principal components were used for the visual mapping which can be seen in Fig.~\ref {fig:TSNE}(a)-(e). Visually, we have not observed any significant difference between full-length embeddings and the top-50 principal components as we have in cluster quality metrics. Nonetheless, We have observed and differentiated the total number of patches (Fig.~\ref{fig:TSNE}(a) \& (d)) extracted from the high-cellular patches (Fig.~\ref{fig:TSNE}(b) \& (e)) which are then used for the atlas. In Fig.~\ref{fig:TSNE}(b) \& (e) we can clearly see the well-separated clusters for 35 classes in the WHO breast atlas when compared with Fig.~\ref{fig:TSNE}(a) \& (d), which can be exploited for search and matching. Fig.~\ref{fig:TSNE}(c) \& (f) finally shows the visualization of the matching results using three test cases as we can see the embeddings of the test images are overlapping or are in close proximity to their correct subtype patches in the atlas.

\subsection*{Conclusion}
In this preliminary work, we examined the idea of a digital atlas indexed by deep embeddings for search and matching in diagnostic histopathology for breast lesions. We also analyzed and visualized the embeddings (generated by KimiaNet, a deep model specialized in histopathology via training on TCGA data) with 35 different primary diagnoses, including rare tumour types. Additionally, we analyzed the 50 principal components of the embeddings and compared them with the full-length feature vectors. Furthermore, we introduced the notion of ``\emph{top-3@top-n}'', a histogram of search results that the top three diagnoses suggestion concept when retrieving top-n matching results to narrow down the broad diagnosis spectrum.

For future work, the atlas concept should be expanded to WSI or regional matching. On the other hand, it could also be used and tested in clinical practice as \textit{virtual second opinion}, which may improve the diagnostic accuracy, increase efficiency, and reduce variability among pathologists.

\bibliography{Bibliographyy}

\section*{Funding}
This collaboration has been initially supported by the Ontario Government, Canada, through an ORF-RE grant. All experiments and analysis work have been supported by an internal grant from Mayo Clinic, Rochester, MN, USA.

\section*{Conflict of Interest Statement}
All authors certify that they have NO affiliations with or involvement in any
organization or entity with any financial interest (such as honoraria; educational grants; participation in speakers’ bureaus;
membership, employment, consultancies, stock ownership, or other equity interest; and expert testimony or patent-licensing
arrangements), or non-financial interest (such as personal or professional relationships, affiliations, knowledge or beliefs) in
the subject matter or materials discussed in this manuscript.

\section*{Disclaimer}
The content of this article represents the personal views of the authors and does not represent the views of the authors’ employers and associated institutions. Where authors are identified as personnel of the International Agency for Research on Cancer/World Health Organization, the authors alone are responsible for the views expressed in this article and they do not necessarily represent the decisions, policy, or views of the International Agency for Research on Cancer/World Health Organization.

\section*{Acknowledgements}
We would like to acknowledge and thanks the department of Anatomical Pathology, Singapore General Hospital, for the digital slides of breast tumours in the WHO classification of breast tumours book. We would also like to acknowledge and thanks Dr Hannah Wen from Memorial Sloan Kettering Cancer Center (MSKCC), New York, United States for contributing the Mucinous cystadenocarcinoma digital slide. Finally, we thank Anne-Sophie Bres
and Mitra Seyedahmad for their administrative assistance. 

\section*{Author contributions statement}

A.S. and H.R.T. conceived the experiments. A.S. conducted the experiments. L.P., P.H.T., I.A.C., and A.M. helped in acquiring the data. R.G. annotated the data. A.S. and H.R.T. analyzed the results.  A.S. wrote the first manuscript draft. H.R.T. rewrote and edited the paper. All authors reviewed the paper.

\end{document}